\documentclass[%
 reprint,
 superscriptaddress,
 amsmath,amssymb,
 aps,
 prl,
]{revtex4-2}

\usepackage{graphicx}
\usepackage{dcolumn}
\usepackage{bm}

\begin{document}

\preprint{APS/123-QED}

\affiliation{Laboratory of Spin Magnetic Resonance, School of Physical Sciences, Anhui Province Key Laboratory of Scientific Instrument Development and Application, University of Science and Technology of China, Hefei 230026, China}
\affiliation{Hefei National Laboratory, University of Science and Technology of China, Hefei 230088, China}
\affiliation{Department of Physics, Laboratory of Computational Materials Physics, Jiangxi Normal University, Nanchang 330022, China}

\title{Reconfigurable Oxide Nanoelectronics by Tip-induced Electron Delocalization}

\author{Chengyuan Huang}
\altaffiliation{These authors contributed equally to this work.}
\affiliation{Laboratory of Spin Magnetic Resonance, School of Physical Sciences, Anhui Province Key Laboratory of Scientific Instrument Development and Application, University of Science and Technology of China, Hefei 230026, China}

\author{Changjian Ma}
\altaffiliation{These authors contributed equally to this work.}
\affiliation{Laboratory of Spin Magnetic Resonance, School of Physical Sciences, Anhui Province Key Laboratory of Scientific Instrument Development and Application, University of Science and Technology of China, Hefei 230026, China}
 
\author{Mengke Ha}%
\altaffiliation{These authors contributed equally to this work.}
\affiliation{Laboratory of Spin Magnetic Resonance, School of Physical Sciences, Anhui Province Key Laboratory of Scientific Instrument Development and Application, University of Science and Technology of China, Hefei 230026, China}

\author{Longbing Shang}
\email{Contact author: lbShang@jxnu.edu.cn}
\affiliation{Department of Physics, Laboratory of Computational Materials Physics, Jiangxi Normal University, Nanchang 330022, China}

\author{Zhenlan Chen}
\affiliation{Laboratory of Spin Magnetic Resonance, School of Physical Sciences, Anhui Province Key Laboratory of Scientific Instrument Development and Application, University of Science and Technology of China, Hefei 230026, China}
\affiliation{Hefei National Laboratory, University of Science and Technology of China, Hefei 230088, China}

\author{Qing Xiao}
\affiliation{Laboratory of Spin Magnetic Resonance, School of Physical Sciences, Anhui Province Key Laboratory of Scientific Instrument Development and Application, University of Science and Technology of China, Hefei 230026, China}

\author{Zhiyuan Qin}
\affiliation{Laboratory of Spin Magnetic Resonance, School of Physical Sciences, Anhui Province Key Laboratory of Scientific Instrument Development and Application, University of Science and Technology of China, Hefei 230026, China}
\affiliation{Hefei National Laboratory, University of Science and Technology of China, Hefei 230088, China}

\author{Danqing Liu}
\author{Haoyuan Wang}
\author{Dawei Qiu}
\author{Qianyi Zhao}
\author{Ziliang Guo}
\author{Yanling Liu}
\author{Dingbang Chen}
\author{Chengxuan Ye}
\author{Zhenhao Li}
\affiliation{Laboratory of Spin Magnetic Resonance, School of Physical Sciences, Anhui Province Key Laboratory of Scientific Instrument Development and Application, University of Science and Technology of China, Hefei 230026, China}

\author{Chang-Kui Duan}
\affiliation{Laboratory of Spin Magnetic Resonance, School of Physical Sciences, Anhui Province Key Laboratory of Scientific Instrument Development and Application, University of Science and Technology of China, Hefei 230026, China}
\affiliation{Hefei National Laboratory, University of Science and Technology of China, Hefei 230088, China}

\author{Guanglei Cheng}
\email{Contact author: glcheng@ustc.edu.cn}
\affiliation{Laboratory of Spin Magnetic Resonance, School of Physical Sciences, Anhui Province Key Laboratory of Scientific Instrument Development and Application, University of Science and Technology of China, Hefei 230026, China}
\affiliation{Hefei National Laboratory, University of Science and Technology of China, Hefei 230088, China}

\date{\today}

\begin{abstract}
Reconfigurable oxide nanoelectronics, enabled by conductive atomic force microscope (cAFM) lithography, have established complex oxide interfaces as a promising platform for quantum engineering that harnesses emergent phenomena for advanced functionalities. However, this cAFM nanofabrication process can only occur in the air, with simultaneous device decay described under the ``water-cycle'' writing mechanism. These restrictions pose ongoing challenges for device optimization in the quantum regime at mK temperatures. Here, we demonstrate a ``waterless'' cAFM lithography approach that is compatible with vacuum and cryogenic environments. Through oxygen vacancy engineering at the LaAlO$_3$/SrTiO$_3$ interface, we have achieved nonvolatile and reconfigurable cAFM control of nanoscale interfacial polaron-electron liquid transition at mK temperatures with an ultrafine line resolution of 0.85 nm. Supported by first-principles calculations and drift-diffusion modeling, we show that tip-controlled oxygen vacancy electromigration plays a key role. This advancement bridges reconfigurable device fabrication and concurrent characterization \textit{in situ} at mK temperatures, and establishes a versatile Hubbard toolbox for engineering programmable quantum phases in correlated oxides. 
\end{abstract}

\maketitle



Precisely placing and removing single electrons on demand in solid-state systems represents an ultimate control in quantum engineering and nanotechnology. Traditional top-down lithography methods involving masking and etching have achieved some success in patterning specific lattices in semiconductors~\cite{hensgens2017quantum,mortemousque2021coherent,borsoi2024shared}. However, they are still limited by the resolution and disorder to engineer a macroscopic coherent phase. Moiré patterns formed by stacking 2D materials provide a cleaner way to create artificial electron lattices and give rise to a number of fascinating emergent phenomena that are absent in individual layers, including superconductivity~\cite{cao2018unconventional}, correlated insulating phases~\cite{cao2018correlated}, and fractional quantum anomalous hall effects~\cite{cai2023signatures,park2023observation,xu2023observation}. However, stacking 2D materials does not enable truly arbitrary electron placement. Bottom-up atom-by-atom fabrication using scanning tunneling microscopy on top of metallic~\cite{gomes2012designer,drost2017topological,slot2017experimental} or silicon substrates~\cite{wang2022experimental,kiczynski2022engineering} offers flexibility for creating nanostructures with atomic precision, particularly intriguing for solid-state quantum simulations. Yet, atom-by-atom fabrication is either time intensive and environmentally sensitive, or not truly reconfigurable, limiting its applications for large-scale or complex architectures.

Correlated oxides, with their strong electron interactions, offer a versatile platform for quantum engineering and studying correlated phenomena~\cite{hwang2012emergent}. At the LaAlO$_3$/SrTiO$_3$ (LAO/STO) oxide interface, the interplay of multiple degrees of freedom gives rise to emergent phenomena such as a high-mobility 2D electron liquid~\cite{ohtomo2004high}, atomic precision metal-insulator transition~\cite{thiel2006tunable}, superconductivity~\cite{caviglia2008electric,richter2013interface}, exotic electron pairing~\cite{cheng2015electron,cheng2016tunable,briggeman2020pascal,nethwewala2023electron}, and magnetism~\cite{ariando2011electronic,bert2011direct,li2011coexistence}. Although nanofabrication in oxides is typically more challenging than in semiconductors and 2D materials, a unique noninvasive lithography technique was developed to reconfigurably control local metal-insulator transition at LAO/STO interface. Through ``writing'' and ``erasing'' using a conductive atomic force microscope (cAFM) tip with sub-10 nm resolution~\cite{cen2008nanoscale,cen2009oxide}, oxide quantum devices like sketched single-electron transistors (SketchSETs)~\cite{cheng2011sketched}, and ballistic electron waveguides~\cite{cheng2018shubnikov,annadi2018quantized} can be conveniently created. Quantum simulations of synthetic spin-orbit coupling and Kronig-Penny potential in 1D superlattices have also been demonstrated~\cite{briggeman2020engineered,briggeman2021one}. The cAFM fabrication-enabled oxide interfaces thus provide a promising platform for implementing correlated quantum nanoelectronics~\cite{coll2019towards}. 

However, a formidable challenge is present in cAFM lithography when device complexity increases. Specifically, devices decay simultaneously during writing in ambient conditions, making parameter optimization extremely difficult. This decay is related to the ``water-cycle'' writing mechanism~\cite{bi2010water}, in which the positively biased cAFM tip locally protonizes LAO/STO sample surface by removing OH$^-$ dissociated from adsorbed water. The underlying LAO/STO interface, initially insulating due to the subcritical LAO thickness of 3 unit cell (uc), becomes conducting under triggered charge transfer~\cite{li2012writing}. Simultaneous decay is induced by the recombination of H$^+$ and OH$^-$ in the air and stops in vacuum. Conversely, a negative bias on the tip removes H$^+$ and completely restores the interface to the insulating state. This process highlights the importance of water. Indeed, writing is unsuccessful in low-humidity environments (typically $<25\%$) or in vacuum. In addition to device decay, it is often hard to predict exact device behaviors at low temperatures during the fabrication process at room temperature. To reach optimal results, devices are frequently warmed up, erased, and rewritten during different temperature cycles, which are very time-consuming. A reconfigurable cAFM nanofabrication operating \textit{in situ} at low temperatures is thus highly desirable. 

Here in this work, we demonstrate a novel method to implement correlated quantum nanoelectronics by reconfigurably localizing and delocalizing single electrons at mK temperatures with ultrafine resolution. Devices are written and measured \textit{in situ} cryogenically, and measurement results are fed back for optimization in subsequent erasing and rewriting procedures. Two challenges were addressed. First, we discovered a new writing mechanism by designing special LAO/STO samples where electrons are localized in the interfacial polarons at subcritical LAO thickness and can be delocalized upon oxygen vacancy ($V_{\text{O}}$) doping controlled by a charged cAFM tip. Such a polaron state has been previously verified by sum-frequency generation spectroscopy~\cite{liu2023nonlinear}. Second, a large scan and ultrastable milliKelvin AFM (mK-AFM) was developed to work in a cryogen-free dilution refrigerator based on a design described elsewhere~\cite{huang2025millikelvinatomicforcemicroscope}. Atomic resolution in the \textit{Z} direction was achieved by minimizing tip-sample vibrations to suppress disorders in fabricated devices. This ``waterless'' cAFM platform thus provides the long-sought control of manipulating single electrons in a solid-state system in a similar fashion to the ultracold atomic lattices for atom manipulations.



\begin{figure}
\centering
\includegraphics[trim={0.5cm 0.3cm 0.2cm 0.5cm}, clip, width=0.5\textwidth]{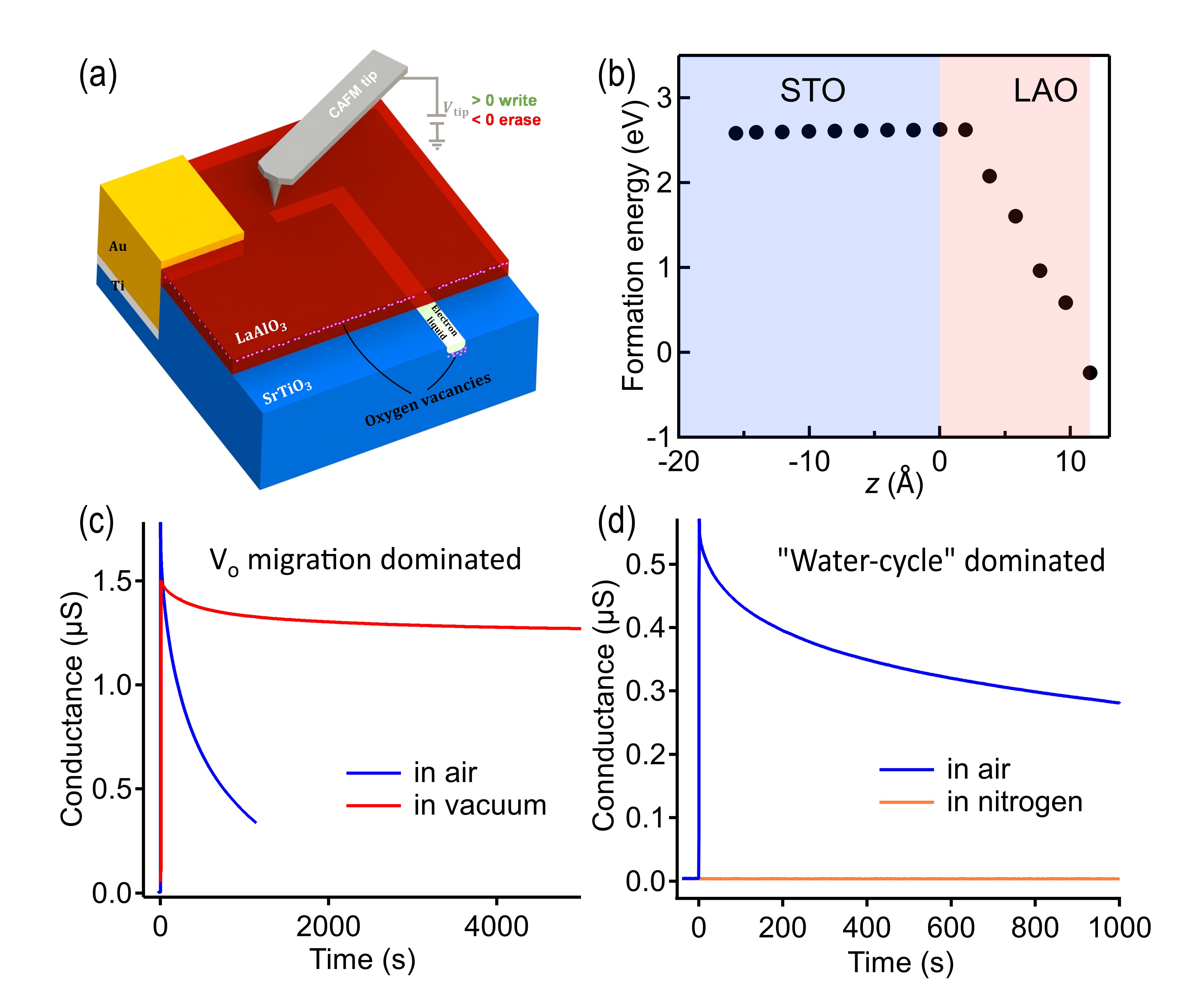}
\caption{cAFM lithography at room temperature. (a) Schematic of sample structure and experiment set-up for ``waterless'' cAFM lithography. The LAO surface layer hosts a programmable $V_{\text{O}}$ reservoir. A sharp AFM tip with different bias voltages locally controls interfacial metal-insulator transition. (b) $V_{\text{O}}$ formation energy profile across the LAO/STO interface, which decreases with increasing LAO thickness. (c) Nanowire writing tests on Sample A in different ambient conditions at room temperature. The slow-decay nanowire written in vacuum is dominated by $V_{\text{O}}$ migration mechanism. (d) Nanowire writing tests on a conventionally grown control sample at room temperature. Writing is unsuccessful in the water-free environment. The fast-decay nanowire written in the air is dominated by the ``water-cycle'' mechanism.}\label{fig1}
\end{figure}

We adopted a modulation doping ($m+n$) uc LAO/STO sample (Supplemental Materials, Sec. \uppercase\expandafter{\romannumeral1}) that features extremely high mobility and previously allows us to observe a time-reversal-symmetry protected transport\cite{mengke2025time}. In this configuration, the interfacial buffer LAO layer ($m\sim 2\text{-}3$) is defect-free with a built-in polarization that tilts the bands, whereas the low-temperature grown capping LAO layer ($n\sim 0\text{-}2$) is filled with $V_{\text{O}}$s that remotely dope the interface. Interface conductivity is controlled by the post-annealing oxygen pressure. It can be tuned insulating even for samples with LAO thickness beyond 4 uc widely reported critical thickness\cite{thiel2006tunable}. These samples were found writable by cAFM in vacuum [Fig. \ref{fig1}(a)] (Supplemental Materials, Sec. \uppercase\expandafter{\romannumeral2}) .

Figure \ref{fig1}(b) shows the calculated $V_{\text{O}}$ formation energy (Supplemental Materials, Sec. \uppercase\expandafter{\romannumeral5}) across the interface, which is minimized at the LAO surface due to the intrinsic polarization field and remains high in STO. In the cubic perovskite structure, oxygen atoms are interconnected through a network of oxygen octahedra, providing open pathways for $V_{\text{O}}$s to migrate under external stimuli\cite{hanzig2016anisotropy,das2017controlled}. Under positive tip biases, surface $V_{\text{O}}$s are driven across the heterointerface into STO, delocalizing electrons trapped in the polarons beneath the tip. After retracting the tip, $V_{\text{O}}$s preferentially stay in STO due to its low ionic mobility, sustaining a conducting electron liquid state. Conversely, negative tip biases drive $V_{\text{O}}$s back to the surface, restoring polaron self-localization. Figure \ref{fig1}(c) shows the cAFM writing tests of nanowires on Sample A (($3+0$) uc LAO/STO) at room temperature. The conductance between two adjacent electrodes is subject to an abrupt increase upon successful nanowire formation. Remarkably, the nanowire can be written in vacuum and exhibits significantly slower decay, retaining 80$\%$ conductance over 12 hours, which conflicts with the ``water-cycle'' mechanism. In comparison, cAFM tests on a conventionally grown control sample (following sample growth details given in refs. \cite{cen2008nanoscale,cen2009oxide,bi2010water,cheng2011sketched,cheng2013anomalous,cheng2015electron,nethwewala2023electron}) suggests the ``water-cycle'' mechanism is dominant [Fig. \ref{fig1}(d)]. The absence of nanowire formation in dry nitrogen conditions is consistent with previous reports\cite{bi2010water}.

\begin{figure}
\centering
\includegraphics[trim={0cm 0.3cm 0cm 0.4cm}, clip, width=0.77\linewidth]{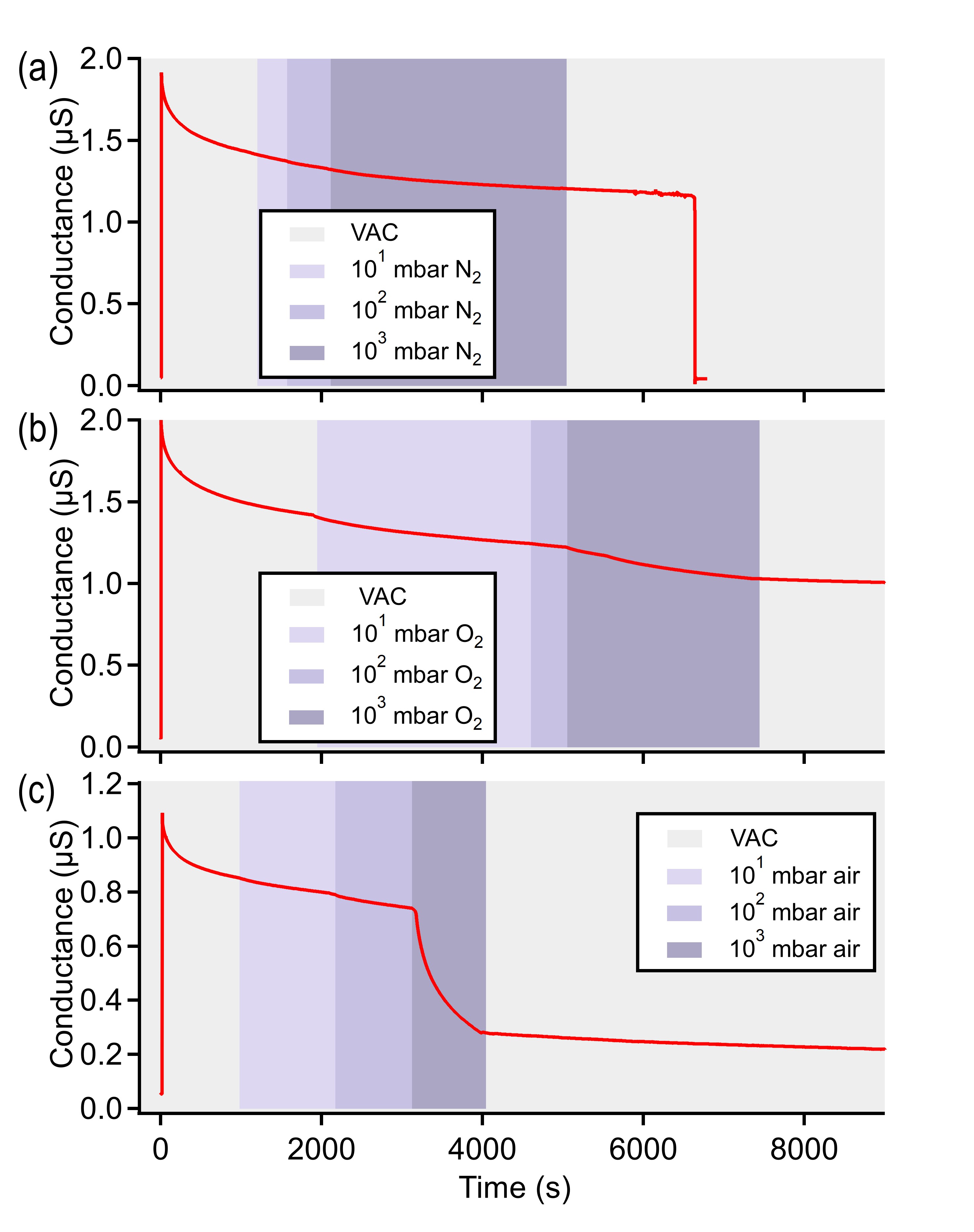}
\caption{Environmental dependence of nanowire decay at room temperature. Nanowires were written on Sample A with 10 V tip bias in vacuum (10$^{-3}$ mbar) at room temperature, then the conductance was monitored under varying environments. (a) Decay in vacuum, followed by dry nitrogen exposure (from 10$^1$ to 10$^3$ mbar), then return to vacuum. The wire was cut at $t\approx6600$ s with -10 V, which resulted in a sharp conductance drop. (b) Decay in vacuum, followed by dry oxygen (or air in (c)) exposure at 10$^1$ to 10$^3$ mbar, then return to vacuum.}
\label{fig2}
\end{figure}

To better understand the decay mechanism of ``waterless'' cAFM lithography, we studied nanowire retention in various ambient environments using a homemade mK-AFM with a vacuum loading chamber. Figure \ref{fig2}(a) shows that the nanowire decay rate remains the same as in the vacuum condition after introducing dry nitrogen, suggesting nitrogen is irrelevant to the decay. Dry oxygen instead has a notable impact on the decay rate [Fig. \ref{fig2}(b)], especially in $10^3$ mbar pressure. This process consistent with surface oxidation in oxygen environment via $\frac{1}{2}\text{O}_2+2e^-+V_{\text{O}}^{2+}\rightarrow\text{O}_{\text{O}}^{\times}$, where $\text{O}_{\text{O}}^{\times}$ denotes a neutral oxygen atom occupying the lattice site\cite{jiang2011mobility}. Namely, the ambient oxygen recombines with the $V_{\text{O}}$s and accelerates the wire decay. Significant impact is observed upon introducing $10^3$ mbar air of 40\% humidity, as shown in Fig. \ref{fig2}(c). This decay has the same origin as in the ``water-cycle'' mechanism. Namely, a proton in the surface water absorbate occupies one $V_{\text{O}}$ and localizes two electrons, which is described by $\text{H}^++2e^-+V_{\text{O}}^{2+}\rightarrow\text{H}_\text{O}^+$, where $\text{H}_\text{O}^+$ represents a proton occupying $V_{\text{O}}$ site. 

\begin{figure}
\centering
\includegraphics[trim={0.5cm 0.3cm 0.2cm 0.5cm}, clip, width=0.5\textwidth]{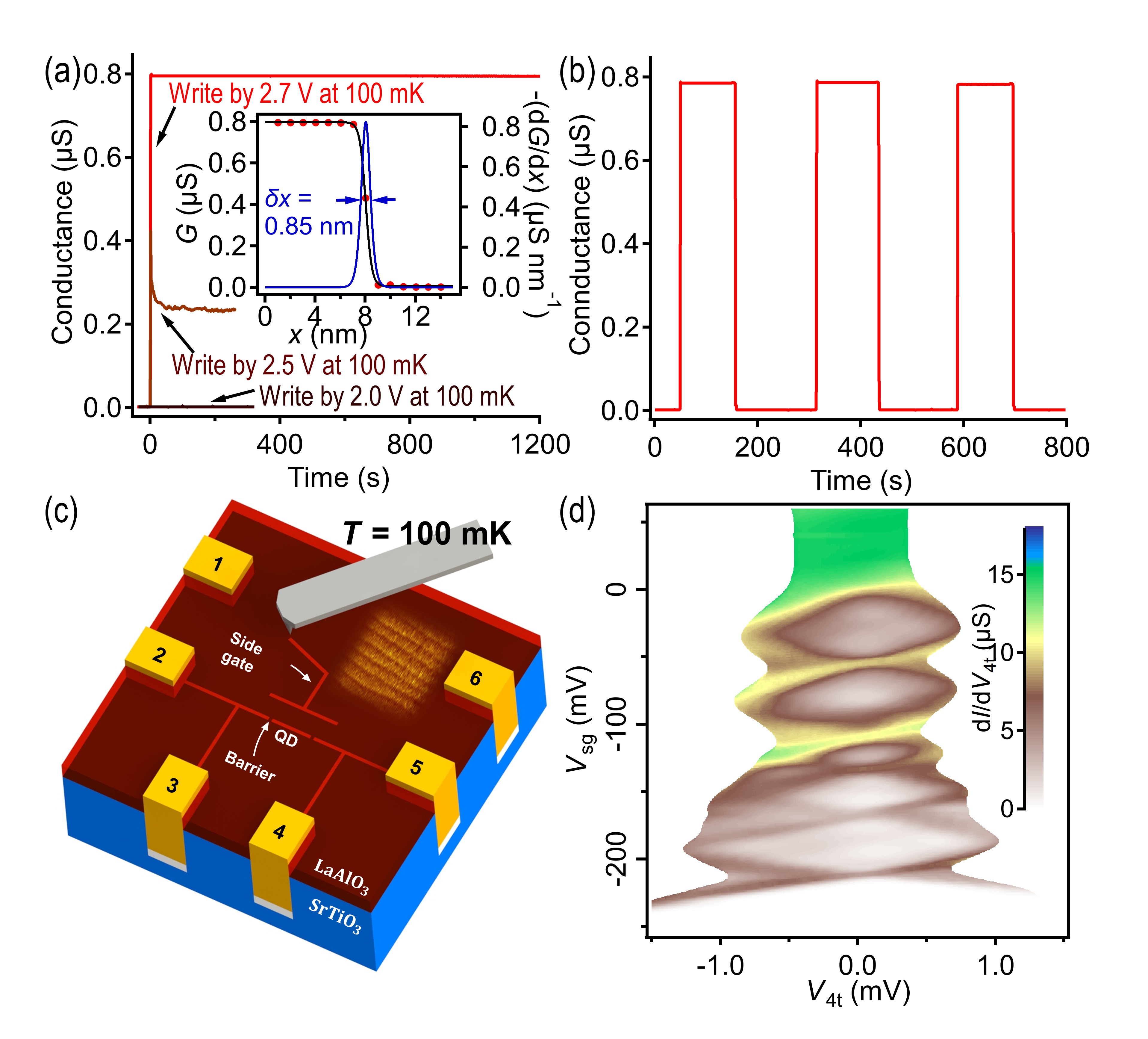}
\caption{Reconfigurable nanofabrication and characterization at 100 mK. (a) Nanowire writing tests at 100 mK on Sample B. 2.0 V tip bias yields no trace, 2.5 V fails to form a stable nanowire, and 2.7 V writes an ultra-fine nanowire showing negligible decay. Red dots in the inset show nanowire conductance measured as a function of the tip position, cutting across the nanowire with -1 V. The sharp conductance drop can be fitted to a profile $G(x)$ (black curve), and the differential conductance $-(\text{d}G(x)/\text{d}x)$ (blue curve) shows a full-width at half-maximum $\delta x = 0.85$ nm. (b) Repeated nanowire writing and erasing at a same position confirms reconfigurability at 100 mK. (c) Schematic of SketchSET fabrication at 100 mK. Inset shows atomic terrace scanned by custom-built mK-AFM. (d) Color-coded $\text{d}I/\text{d}V_{\text{4t}}$ map as a function of $V_{\text{4t}}$ and $V_{\text{sg}}$ for an SketchSET written on Sample C with a nanowire quantum dot (QD) length of 100 nm at 100 mK.}\label{fig3}
\end{figure}
 
\begin{figure*}
\centering
\includegraphics[trim={0cm 0.4cm 0cm 0.4cm}, clip, width=1\textwidth]{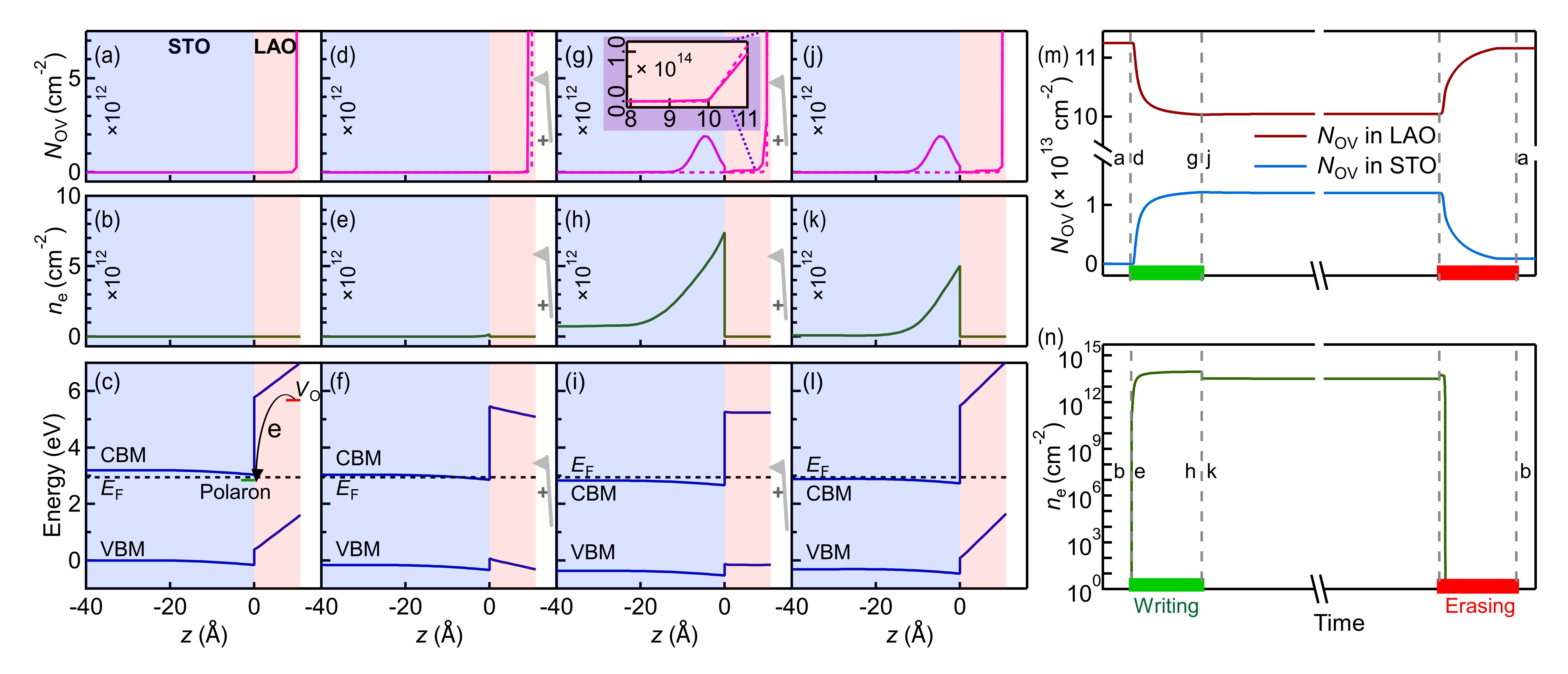}
\caption{Drift-diffusion model of oxygen vacancy migration. (a-c) $N_{\text{OV}}$, $n_{\text{e}}$, and band alignment before writing. In (c),the green line denotes charge transition energy level of polarons, whereas the red line denotes $V_{\text{O}}$ charge transition energy level. (d-f) $N_{\text{OV}}$, $n_{\text{e}}$, and band alignment at the onset of applying a positive bias (solid curves). (g-i) $N_{\text{OV}}$, $n_{\text{e}}$, and band alignment during the writing process (solid curves). The inset of (g) shows a zoomed-in profile near the surface. (j-l) $N_{\text{OV}}$, $n_{\text{e}}$, and band alignment after the writing process (solid curves). The purple dashed curves in (d, g, j) show $N_{\text{OV}}$ before writing for comparison. (m) Total $N_{\text{OV}}$ in LAO layer (red curve) and STO layer (blue curve) as a function of time. (n) Total $n_{\text{e}}$ in STO layer as a function of time. In (m) and (n), a positive voltage is applied between the first two gray dashed lines, and a negative voltage between the last two gray dashed lines. The labels in (m, n) correspond to the stages of $N_{\text{OV}}$ and $n_{\text{e}}$ profile in (a-l).}\label{fig4}
\end{figure*}

The accelerated decay in the air compared to dry oxygen is attributed to the lower adsorption energy and reduced reaction barrier of H$^+$ relative to $\text{O}_2$. Upon re-evacuating, the nanowires in Figs. \ref{fig2}(b, c) resume their slow decay, confirming environmental impacts. The slow decay even in vacuum stems from thermally activated $V_{\text{O}}$ mobility at room temperature with diffusion coefficient $D=\mu k_{\text{B}}T$ (where $\mu$ is the vacancy mobility and $k_{\text{B}}$ is the Boltzmann constant). This process is effectively suppressed at low temperatures (as confirmed by both experiment [Fig. \ref{fig3}(a)] and calculation [Fig. \ref{fig4}(n)]), where the enhanced relative permittivity of STO further elevates the diffusion barrier. These results underscore the critical role of $V_{\text{O}}$ migration in ``waterless'' cAFM lithography, as evidenced by the successful writing on $V_{\text{O}}$ modulation doping ($m+n$) uc samples in vacuum and the pronounced sensitivity of nanowire decay to temperature and ambient environments. A summary of tip bias effects and their respective contributions is provided in Supplemental Materials, Sec. \uppercase\expandafter{\romannumeral6}.


Next, we demonstrate a concept of reconfigurable nanofabrication and characterization \textit{in situ} at 100 mK, a milestone in quantum device nanofabrication. Each sample exhibits a threshold tip bias that is strongly dependent on its quality. The ($m+n$) uc growth method allows precise control of $V_{\text{O}}$ density in LAO surface layer, thereby enabling a lower threshold tip bias for nanowire writing compared to conventionally grown samples, which typically require 10-15 V\cite{cheng2013anomalous}. As shown in Fig. \ref{fig3}(a), a nanowire can be written at Sample B (($2+2$) uc LAO/STO) with a threshold tip bias of about 2.7 V at 100 mK, resulting in an abrupt two-terminal conductance jump from 2 nS to 0.8 $\mu$S, with no noticeable decay. Reversibly, the wire was subsequently cut with -1 V tip bias, restoring the insulating state. The inset of Fig. \ref{fig3}(a) shows the abrupt conductance drop while cutting at a low speed (10 nm/s). The corresponding differential conductance profile can be used to extract the wire width (Supplemental Materials, Sec. \uppercase\expandafter{\romannumeral3}) and yields an ultrafine width of 0.85 nm. This resolution surpasses previously reported minimal feature sizes of devices fabricated in the air in LAO/STO\cite{cen2008nanoscale,eom2021electronically} and LAO/KTO\cite{yu2022nanoscale} samples. The enhancement is likely due to two factors: (i) a smaller tip-sample contact area by eliminating the water bridge mediating tip-sample interaction in the air. (ii) negligible thermal activation to ensure only $V_{\text{O}}$s under the contact area are activated by the electric field. In addition, we emphasize the ultra-stability of our homemade mK-AFM, as evidenced by resolving the atomic terraces in vibration-intensive dilution refrigerator [inset of Fig. \ref{fig3}(c)], is vital to a successful writing with high quality. 

To highlight reconfigurability, the wire was erased by high-resolution raster scanning in a 1 $\mu$m $\times$ 1 $\mu$m area with -1 V tip bias, and a new wire was written at the same position of the sample canvas in a similar fashion repeatedly [Fig. \ref{fig3}(b)]. 

Complex quantum devices can be readily created using this method. Figure \ref{fig3}(c) illustrates the \textit{in situ} fabrication of a SketchSET on Sample C (($3+2$) uc LAO/STO) at 100 mK following a protocol described previously\cite{cheng2011sketched,cheng2015electron}. The critical step is barrier formation via gentle cutting across a nanowire. It usually takes multiple cool-downs and cryogenic measurements to determine the appropriate barrier parameters if fabricated in the air. Here, this process can be greatly simplified due to the reconfigurable nature and \textit{in situ} measurement at 100 mK. As a result, this SketchSET comprises a 100 nm nanowire bounded by two tunneling barriers, with a side gate 500 nm away. The differential conductance ($\text{d}I/\text{d}V_{\text{4t}}$) as a function of four-terminal voltage $V_{\text{4t}}$ and side gate voltage $V_{\text{sg}}$ is shown in Fig. \ref{fig3}(d), which shows the characteristic Coulomb diamonds signifying single electron tunneling. Superconductivity is not observed, likely due to the high electron temperature ($T_{\text{e}}\approx 340$ mK) in the mK-AFM. 


The $V_{\text{O}}$ migration across the LAO/STO interface can be modeled by drift-diffusion equations (Supplemental Materials, Sec. \uppercase\expandafter{\romannumeral4}). Initially, the surface-concentrated $V_{\text{O}}$ distribution was tuned to a density of $\sim$0.18 uc$^{-2}$ during sample growth. This value corresponds to a critical insulating state, where the Fermi energy $E_{\text{F}}$ is just below the conduction band minimum (CBM) of STO and $\sim$1.3 eV above the valence band maximum (VBM) of LAO top layer [Figs. \ref{fig4}(a-c)]. This alignment is estimated by the equilibrium between the polarization field in LAO and $V_{\text{O}}$ formation in LAO top layer where the $V_{\text{O}}$ formation energy is close to zero. Electrons released during $V_{\text{O}}$ formation in sample growth are self-trapped via Ti$^{4+}$ reduction in STO to form polarons\cite{franchini2021polarons}, and also by relatively sparse deep-level defects like cation antisites and cation vacancies.

Upon applying a positive tip bias, electronic bands bend downward and cause $E_{\text{F}}$ to slightly pass CBM [Figs. \ref{fig4}(d-f)]. In the meantime, $V_{\text{O}}$s migrate from the LAO surface layer into the STO under the electric field [Fig. \ref{fig4}(g)]. Positively charged $V_{\text{O}}$s in STO progressively depress the conductance band below $E_{\text{F}}$ [Fig. \ref{fig4}(i)], facilitating the formation of electron liquid at the interface [Figs. \ref{fig4}(e, h)]. The extent of $V_{\text{O}}$ migration depends on the magnitude of the applied bias and writing speed, with spatial resolution determined by the tip apex geometry. After the removal of the positive bias, stabilized $V_{\text{O}}$s in STO [Fig. \ref{fig4}(j)] maintain CBM below $E_{\text{F}}$ [Fig. \ref{fig4}(l)], preserving the interfacial conductivity in a non-volatile fashion [Fig. \ref{fig4}(k)]. Figures \ref{fig4}(m, n) show the dynamics of $V_{\text{O}}$ density $N_{\text{OV}}$ and free electron density $n_{\text{e}}$ during ``waterless'' cAFM lithography at 3 K, which exhibits negligible decay after bias removal. The evolution of $N_{\text{OV}}$ and $n_{\text{e}}$ during erasing inversely repeats the writing process.

In conclusion, we have demonstrated reconfigurable oxide electronics at precisely engineered LAO/STO oxide interfaces, fully compatible with vacuum and cryogenic operations. The underlying mechanism exploits tip-controlled $V_{\text{O}}$ electromigration to locally trigger reversible metal-insulator transitions. Unlike conventional atom-by-atom manipulation using STM, our method enables arbitrary placement and removal of single electrons with ultrahigh precision at mK temperatures. Our work offers a new strategy for creating programmable 1D/2D electron lattices, providing a foundational toolkit for ``Hubbard toolbox'' engineering in correlated oxides.

 \textit{Acknowledgments}—This work was supported by the National Key Research and Development Program of China (2024YFA1409500), the CAS Project for Young Scientists in Basic Research (YSBR-100), the Fundamental Research Funds for Central Universities (KY2030000160 and WK3540000003), and the National Natural Science Foundation of China (Grant No. 12164019, 12364026, 12464029, 12174162).

\bibliography{Reference}

\end{document}